\documentclass[twocolumn,showpacs,preprintnumbers,amsmath,prl,amssymb,superscriptaddress]{revtex4}

\usepackage{graphicx}
\usepackage{dcolumn}
\usepackage{bm}

\def\th232{\rm{ ^{232} Th }}
\def\ur238{\rm{ ^{238} U }}
\def\u235{\rm{ ^{235} U }}
\def\p239{\rm{ ^{239} Pu }}
\def\pu241{\rm{ ^{241} Pu }}
\def\chr51{\rm{ ^{51} Cr }}
\def\fe55{\rm{ ^{55} Fe }}
\def\ni59{\rm{ ^{59} Ni }}
\def\ynu{\rm{Y_{\nu}}}
\def\ynubar{\rm{Y_{\bar{\nu}}}}
\def\ung{\rm{^{238}U ( n , \gamma )}}

\def\nuebar{\rm{\bar{\nu_e}}}
\def\nue{\rm{\nu_e}}
\def\keff{\rm{K_{eff}}}
\def\sigmang{\rm{\sigma _{n \gamma}}}
\def\nufluxunit{\rm{cm^{-2} s^{-1}}}
\def\s2tw{\rm{ sin ^2 \theta _W }}
\def\munu{\rm{\mu_{\nu}}}

\def\mub{\rm{\mu_B}}
\def\enu{\rm{E_{\nu}}}

\def\offprime{\it{OFF^{\ast}}}

\begin{document}

\preprint{AS-TEXONO/05-03}

\title{
Production of Electron Neutrinos at Nuclear
Power Reactors \\
and the Prospects for Neutrino Physics
}

%
\newcommand{\as}{Institute of Physics, Academia Sinica, Taipei 115, Taiwan.} 
\newcommand{\ntu}{Department of Physics, National Taiwan University,
Taipei 106, Taiwan.}
\newcommand{\ihep}{Institute of High Energy Physics, 
Chinese Academy of Science, Beijing 100039, China.}
\newcommand{\thu}{Department of Engineering Physics, Tsing Hua University,
Beijing 100084, China.}
\newcommand{\umd}{Department of Physics, University of Maryland,
College Park MD 20742, U.S.A.}
\newcommand{\ks}{Kuo-Sheng Nuclear Power Station, 
Taiwan Power Company, Kuo-Sheng 207, Taiwan.}
\newcommand{\iner}{Institute of Nuclear Energy Research, 
Lung-Tan 325, Taiwan.}
\newcommand{\ckit}{Department of Management Information Systems,
Chung Kuo Institute of Technology, Hsin-Chu 303, Taiwan.}
\newcommand{\ciae}{Department of Nuclear Physics, 
Institute of Atomic Energy, Beijing 102413, China.}
\newcommand{\paris}{ LPNHE, Universit\'{e} de Paris VII,
Paris 75252, France}
\newcommand{\corr}{htwong@phys.sinica.edu.tw}

\affiliation{ \ciae }
\affiliation{ \as }
\affiliation{ \umd }
\affiliation{ \ihep }
\affiliation{ \thu }
\affiliation{ \paris }

\author{ B.~Xin } \affiliation{ \ciae } \affiliation{ \as }
\author{ H.T.~Wong } \altaffiliation[Corresponding Author: ]{ \corr } \affiliation{ \as }
\author{ C.Y.~Chang } \affiliation{ \as } \affiliation{ \umd }
\author{ C.P.~Chen }  \affiliation{ \as } 
\author{ H.B.~Li }  \affiliation{ \as } 
\author{ J.~Li }  \affiliation{ \ihep } \affiliation{ \thu }
\author{ F.S.~Lee } \affiliation{ \as }
\author{ S.T.~Lin } \affiliation{ \as } 
\author{ V.~Singh } \affiliation{ \as }
\author{ F.~Vannucci } \affiliation{ \paris }
\author{ S.C.~Wu } \affiliation{ \as }
\author{ Q.~Yue } \affiliation{ \ihep } \affiliation{ \thu }
\author{ Z.Y.~Zhou } \affiliation{ \ciae }

\collaboration{TEXONO Collaboration}

\noaffiliation


\date{\today}

\begin{abstract}

High flux of 
electron neutrinos($\nue$) is produced at 
nuclear power reactors  
through the decays of 
nuclei activated by neutron capture.
Realistic simulation studies on the neutron
transport and capture at the reactor core were performed.
The production of
$\chr51$ and $\fe55$ give rise to 
mono-energetic $\nue$'s at 
Q-values of 753~keV and 231~keV
and fluxes of $8.3 \times 10^{-4}$ and 
$3.0 \times 10^{-4}$ $\nue$/fission, respectively.
Using data from a germanium detector 
at the Kuo-Sheng Power Plant, 
we derived direct limits on the
$\nue$ magnetic moment and the radiative
lifetime of 
$\mu_{\nu} < 1.3 \times 10^{-8} ~ \mub$ and 
$\rm{ \tau_{\nu} / m_{\nu} > 0.11~s / eV}$ 
at 90\% confidence
level (CL), respectively.
Indirect bounds on 
$\rm{ \tau_{\nu}  / m_{\nu}^3}$
were also inferred.
The $\nue$-flux can be enhanced
by loading selected isotopes to the
reactor core, and the potential
applications and achievable statistical
accuracies were examined. 
These include accurate cross-section
measurements, studies of 
mixing angle $\theta_{13}$
and monitoring of plutonium
production.

\end{abstract}


\maketitle

\section{I. Introduction}

Results from recent neutrino experiments
provide strong evidence for neutrino oscillations
due to finite neutrino masses and
mixings\cite{pdg,nu04}.
Their physical origin and experimental consequences
are not fully understood.
Studies on neutrino properties
and interactions can shed light on these
fundamental questions and 
constrain theoretical models necessary for the
interpretation of future precision data.
It is therefore motivated to explore alternative neutrino
sources and new neutrino detection channels.

The theme of this paper is to study the production
of electron neutrinos($\nue$) from 
nuclear power reactors.
Fluxes derived from the 
``Standard Reactor''
configuration were used to 
obtain direct limits on the neutrino properties 
from data taken at the Kuo-Sheng Power Plant. 
The hypothetical ``Loaded Reactor'' 
scenario was also
studied, where selected
materials were inserted to the core to 
substantially enhance
the $\nue$-flux.
The detection channels
and the achievable physics potentials 
in ideal experiments were investigated.

\section{II. Standard Power Reactor}

\subsection{A. Evaluation of Electron Neutrino Fluxes}

Production of electron anti-neutrinos($\nuebar$) 
due to $\beta$-decays of fission products at 
power reactors is a well-studied process.
There are standard parametrizations for the
reactor $\nuebar$ spectra\cite{vogelengel}.
The typical fission rate at 
the reactor core with a thermal power of 
$\rm{P_{th}}$ in GW is 
$\rm{0.3 \times 10^{20} ~P_{th} ~ s^{-1} }$,
while an average about 6 $\nuebar$/fission are emitted.
The modeling of the $\nuebar$ energy spectra above 3~MeV 
is consistent
with measurements at the $<$5\% level\cite{bugey3},
while the low energy portion is
subjected to much bigger uncertainties\cite{lernu}.
In a realistically achievable setting 
at a location 10~m from a core with
$\rm{P_{th}}$=4.5~GW, the $\nuebar$-flux
is $6.4 \times 10^{13}~\nufluxunit$.

Nuclear reactors also produce $\nue$ via (a) electron 
capture or inverse beta decay of the
fission products and 
(b) neutron activation on the fuel rods and the construction
materials at the reactor core. 
There were unpublished studies\cite{rnu70} 
on the reactor $\nue$-fluxes 
from early reactor experiments, 
indicating that they
would not contribute to the
background in
the measurements with $\nuebar$.
We extended these studies
with realistic simulations,
and focused 
on the potentials
of using them as sources 
to study neutrino physics.

Primary fission daughters are predominantly neutron-rich 
and go through $\beta^-$-decays
to reach stability. Direct feeding to isotopes which decay
by $\beta ^+$-emissions  or electron capture(EC)
is extremely weak, at the $\sim$$10^{-8}$/fission 
level\cite{database}.  
The leading components for
the $\u235$ and $\p239$ fissions with
relative contributions $\rm{r_f}$,
fission yields $\rm{Y_f}$ and
branching ratio BR for $\nue$-emissions
are shown in Table~\ref{fissionyield}a.
The average $\nue$-yield per fission
$\ynu$ is therefore
$\rm{\ynu = r_f \cdot Y_f \cdot BR}$.
In addition, stable fission products 
can undergo (n,$\gamma$) capture
to unstable states which
decay by $\nue$-emissions.
The equilibrium yield of the major
components\cite{database} 
are shown in Table~\ref{fissionyield}b.
The leading contribution 
is from $\rm{^{103}Rh ( n, \gamma ) ^{104}Rh}$,
where
the yield summed over all four
fissile isotopes is 
$\ynu = 2.1 \times 10^{-4} ~ \nue$/fission.
However, under realistic settings in
reactor operation,
the time to reach equilibrium is of the order of
10~years, such that the contribution to
$\nue$-emissions
by this channel is  also small
($\ynu  \sim 10^{-5}$).

\begin{table}
\caption{\label{fissionyield}
The leading $\nue$-yields per fission
($\rm{\ynu = r_f \cdot Y_f \cdot  BR}$) from
(a) direct feeding of daughters (Z,N)
and (b) neutron capture on stable
isotope (Z,N-1) at equilibrium conditions.
}
\begin{ruledtabular}
(a)
\begin{tabular}{lccccc}
Series & (Z,N) & $\rm{Y_f (Z,N)}$ &
$\rm{Q(MeV)}$ & BR(\%) & $\ynu$ \\ \hline
$\u235$ & $^{86}$Rb  & 1.4e$^{-5}$  &
0.53 & 0.005  &  4.3e$^{-10}$ \\
($\rm{r_f = 0.62}$) & $^{87}$Sr & $<$1e$^{-5}$ &
0.2 & 0.3 & $<$1.9e$^{-8}$ \\
& $^{104}$Rh & 7e$^{-8}$ &
1.15 & 0.45 & 2.0e$^{-10}$ \\
& $^{128}$I & 1.2e$^{-8}$ &
1.26 & 6 & 4.3e$^{-10}$ \\ \hline
$\p239$ & $^{128}$I & 1.7e$^{-6}$ &
1.26 & 6 & 2.6e$^{-8}$ \\
($\rm{r_f = 0.26}$) & $^{110}$Ag & 1.3e$^{-5}$ &
0.88 & 0.3 & 1.0e$^{-8}$ \\
\end{tabular}
\vspace*{0.2cm}
(b)
\begin{tabular}{lcccccc}
Series & (Z,N) & $\rm{Y_f}$(A-1) &
$\sigmang$(b) &
$\rm{Q(MeV)}$ & BR(\%) & $\ynu$ \\ \hline
$\u235$ & $^{104}$Rh & 3.2e$^{-2}$ & 146 &
1.15 & 0.45 & 9.0e$^{-5}$  \\
($\rm{r_f = 0.62}$) & $^{128}$I & 1.2e$^{-3}$ & 6.2 &
1.26 & 6 & 4.3e$^{-5}$ \\
& $^{122}$Sb & 1.2e$^{-4}$ & 6.2 &
1.62 & 2.2 & 1.6e$^{-6}$ \\
& $^{110}$Ag & 3e$^{-4}$ & 89 &
0.88 & 0.3 & 5.6e$^{-7}$ \\ \hline
$\p239$ & $^{128}$I & 5.2e$^{-3}$ & 6.2 &
1.26 & 6 & 8.2e$^{-5}$ \\
($\rm{r_f = 0.26}$) & $^{104}$Rh & 6.8e$^{-2}$ & 146 &
1.15 & 0.45 & 8.0e$^{-5}$  \\
& $^{110}$Ag & 1.1e$^{-2}$ & 89 &
0.88 & 0.3 & 8.9e$^{-6}$ \\
& $^{122}$Sb & 4.3e$^{-4}$ & 6.2 &
1.62 & 2.2 & 2.6e$^{-6}$
\end{tabular}
\end{ruledtabular}
\end{table}

\begin{figure}[htb]
\includegraphics[width=8cm]{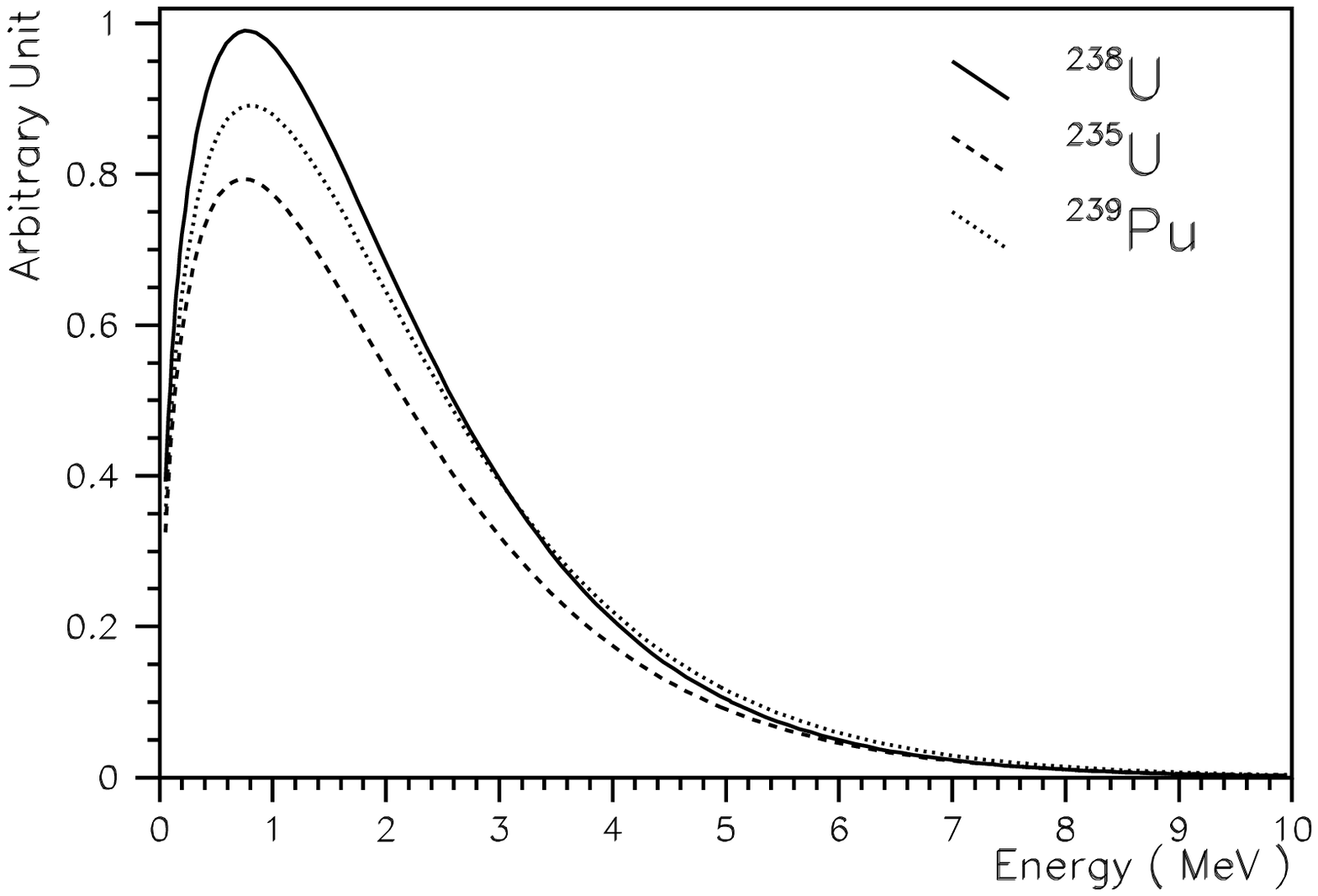}
\caption{
The energy spectra for emitted neutrons
from the fissile isotopes $\u235$, $\ur238$
and $\p239$. The spectra for $\pu241$
is approximated to be that from $\p239$.
}
\label{watt}
\end{figure}

A complete ``MCNP''
neutron transport simulation\cite{mcnp}
was performed to study the effect of neutron
capture on the reactor core materials, which
include the fuel elements, cooling water,
control rod structures and 
construction materials.
While the layout is generic for most nuclear power
reactors, the exact dimensions and material compositions
were derived from the $\rm{P_{th}}$=2.9~GW
Core\#1 of the Kuo-Sheng(KS) Nuclear
Power Station in Taiwan, where a neutrino laboratory\cite{texono}
has been built. 
The reactor core materials and their mass 
compositions are summarized
in Table~\ref{mcnpinput}a.
A homogeneous distribution of these materials
inside a stainless 
steel containment vessel of 
inner radius 225~cm, height 2750~cm, and 
thickness 22~cm was adopted.
This approximation is commonly used and has been
demonstrated to be valid in reactor design 
studies\cite{reactordesign}.
Standard parametrizations of the
``Watt'' fission neutron 
spectra\cite{mcnp,reactordesign}
were adopted as input:    
\begin{equation}
\rm{
\phi_n \propto  exp(-E/a) ~ sinh( \sqrt{bE} ) 
}
\end{equation}
where (a,b) depend on the fission elements.
The emitted neutron spectra for the 
fissile isotopes are depicted in Figure~\ref{watt}.
There are on average 2.5 neutrons
generated per fission with energy
distribution peaked at $\sim$1~MeV. 
The neutrons are scattered
in the core and eventually absorbed by either  
the (n,fission) processes with the fuel elements or 
the (n,$\gamma$) or other interactions with the core materials.

\begin{table}
\caption{\label{mcnpinput}
The compositions for 
(a) the construction materials
inside the reactor core 
and of the containment
vessel
used in the neutron capture studies, and
(b) the three isotopes responsible
for $\nue$-emissions.
}
\begin{ruledtabular}
(a)
\begin{tabular}{l l c}
Functions & Materials & Weight (kg) \\ \hline
\underline{UO$_2$ Fuel Elements}: & Total & 110000 \\
$~~$Fission Isotopes: & $\u235$ & 1376 \\
& $\ur238$ & 98688 \\
& $\p239$ & 431 \\
& $\pu241$ & 84 \\ \hline  
\underline{Non-Fuel Materials inside} & & \\
$~$ \underline{Containment Vessel}: & Total & 125000 \\
$~~$ Fuel Container & Zr-Alloy & 67500 \\ 
$~~$ Cooling & Water & 42500 \\ 
$~~$ Control Rod Assembly$^\dagger$: & $\rm{B_4 C}$ & 479 \\
& Stainless Steel & 14100 \\ \hline 
\underline{Containment Vessel}: & Stainless Steel & 910000 \\ 
\end{tabular}
$^\dagger$ at complete insertion. \hfill
\vspace*{0.2cm} 

(b)
\begin{tabular}{l c c c}
Materials & \multicolumn{3}{c}{Compositions (\%)}\\ 
& $^{50}$Cr & $^{54}$Fe & $^{58}$Ni \\ \hline
Stainless Steel SUS304 & 0.95 &  4.2 & 6.3 \\
Zr-2 alloy & 0.005 & 0.006 & 0.034 \\
\end{tabular}
\end{ruledtabular}
\end{table}

\begin{figure}[htb]
\includegraphics[width=8cm]{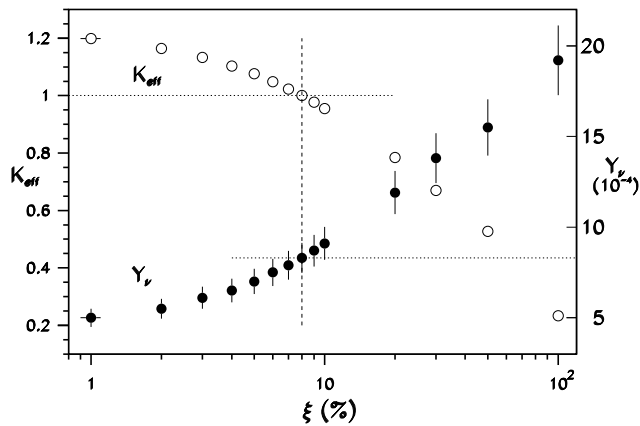}
\caption{
The variations of the  $\keff$ parameter and
the neutrino yield $\ynu$ from $\chr51$, as
functions of control rod fraction  $\xi$.
}
\label{xikeff}
\end{figure}

An important constraint is that 
an equilibrium chain reaction must be 
sustained to provide stable power generation. 
This is achieved by regulating the fraction
of the control rod assembly($\xi$) inserted into
the fuel bundles.
This constraint is parametrized
by $\keff$ defined
as the ratio of neutron-induced fission 
to starting fission rates.
The variation of $\keff$ versus $\xi$ is
depicted in Figure~\ref{xikeff}.
The equilibrium conditions 
require $\keff$=1.0, 
and the distributions of the
per-fission neutron
capture yield($\rm{Y_n}$) 
are given in Table~\ref{mcnpresults}.
Only 0.35\% of the neutrons
escape from the containment vessel,
justifying that detailed treatment
exterior to the vessel is not necessary.

\begin{table}

\caption{\label{mcnpresults}
The neutron capture yields $\rm{Y_n}$ 
of the major channels 
at $\keff$=1.
The average number of neutron emitted
per fission is $\rm{\Sigma Y_n}$=2.5.
}
\begin{ruledtabular}
\begin{tabular}{lccc}
Channel & Isotope & Weight (kg) &  $\rm{Y_n}$ \\ \hline
(n,fission) on & $\ur238$ & 98688  & 0.057 \\ 
fuel element  & $\u235$ & 1376 & 0.62 \\
& $\p239$ & 431 & 0.26 \\
& $\pu241$ & 84 & 0.068 \\ \hline
& \multicolumn{2}{c}{ $\rm{\Sigma Y_n (fission) = }$ }
& 1.0 \\ \hline \hline 
(n,$\gamma$) at &  $\ur238$ & 98688 & 0.59 \\
Core Region 
& Water & 42519 & 0.25 \\
& $\rm{^{10}B}$ & 5.4 & 0.28 \\
&  $^{50}$Cr & 9.0 & 0.00067 \\
&  $^{54}$Fe & 26.6 & 0.00018  \\
&  $^{58}$Ni & 57.6 & 0.0010 \\ \hline
(n,$\gamma$) at 
&  $^{50}$Cr & 8650 & 0.00016 \\
Stainless Steel 
&  $^{54}$Fe & 38200 &  0.00012  \\
Containment Vessel 
&  $^{58}$Ni & 57300 & 0.00026 \\ \hline
\multicolumn{3}{l}{Other capture channels:} & \\
\multicolumn{3}{l}{~~[mainly (n,$\gamma$) on other isotopes]} 
& 0.37 \\ \hline
\multicolumn{3}{l}{External to Containment Vessel} 
&  0.009 \\ \hline
& \multicolumn{2}{c}{ $\rm{\Sigma Y_n (total) = }$ }
& 2.5 
\end{tabular}
\end{ruledtabular}
\end{table}

Stainless steel ``SUS304''
and ``Zr-2 alloy'' are the
typical construction materials at reactor
cores, used in 
the containment vessel and control rod assembly,  
as well as in fuel element containers, respectively.
These materials contain 
$^{50}$Cr, $^{54}$Fe and $^{58}$Ni
with compositions given in Table~\ref{mcnpinput}b.
Upon activation by the (n,$\gamma$) reactions,
these isotopes 
produce $\chr51$, $\fe55$ and $\ni59$ that
will subsequently decay by EC 
and $\nue$-emissions.
Their properties (isotopic abundance IA,
(n,$\gamma$) cross-sections $\sigmang$, 
half-life $\rm{\tau _{\frac{1}{2}}}$, 
EC Q-value and branching ratio BR) 
and $\nue$-yield ($\ynu$=$\rm{Y_n} \cdot$BR) 
are given in Table~\ref{ncapture}. 
The variation of $\ynu$ in
$\chr51$ with $\xi$ is
displayed in Figure~\ref{xikeff}. 
The half-life of $\ni59$ 
is too long and thus not relevant for 
$\nue$-emissions.
The dominant reactor $\nue$ sources 
are therefore $\chr51$ and $\fe55$,
with total yields of 
$\ynu = 8.3 \times 10^{-4}$
and $3.0  \times 10^{-4}$ $\nue$/fission,
implying $\nue$-fluxes of
$\rm{7.5 \times 10^{16} ~ s^{-1}}$ 
and $\rm{2.7 \times 10^{16} ~ s^{-1}}$,
respectively,
at a 2.9~GW reactor.
The total strength corresponds to a 2.7~MCi source.

\begin{table}
\caption{\label{ncapture}
The $\nue$ sources and their yields
$\rm{Y_n}$, $\ynu$(both in $10^{-4}$)
at the reactor core.
}
\begin{tabular}{lcccccccc} 
\hline \hline
Isotope & IA(\%) & $\sigmang$(b)
& $\tau _ {\frac{1}{2}}$ & Q(keV) & BR(\%) &
$\rm{Y_n}$ & $\ynu$ \\ \hline
$^{103}$Rh & 4.6$^{\dagger}$ & 146 & 41.8~s & 1145 & 0.45
& 30$^{\ddagger}$  & 0.14$^{\ddagger}$ \\
$^{50}$Cr & 4.35 & 15.8 & 27.7~d & 753 & 100 & 8.3 & 8.3 \\
$^{54}$Fe & 5.85 & 2.3 & 2.73~y &  231 & 100 & 3.0 & 3.0 \\
$^{58}$Ni & 68.1 & 4.6 & 7.6e$^4$~y & 1073 & 100 & 13.0 & $-$ \\ \hline \hline
\multicolumn{8}{l}{$^{\dagger}$ fission yield} \\
\multicolumn{8}{l}{$^{\ddagger}$ averaged over 18 months reactor period}\\
\end{tabular}
\end{table}

\begin{figure}[htb]
\includegraphics[width=8.0cm]{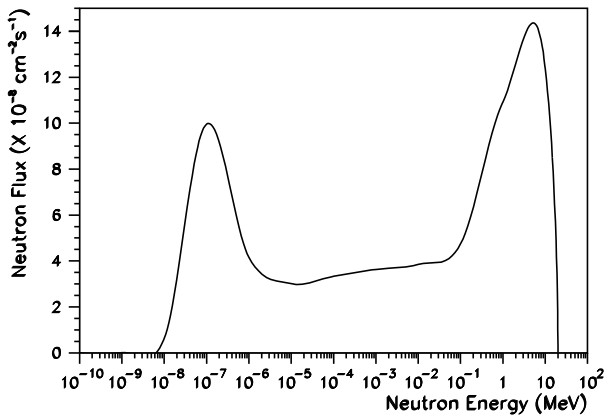}
\caption{
Energy spectrum of neutron at the reactor core,
derived from MCNP simulations. 
}
\label{coreneutron}
\end{figure}

To demonstrate the 
validity of the simulation procedures and results,
a series of cross-checks were made.
As depicted in Figure~\ref{xikeff}, 
the control rod fraction is $\xi$=8\% at 
critical condition $\keff$=1. 
The relative fission yields of the
four fissile elements are
given in Table~\ref{mcnpresults}. 
The neutron energy spectrum averaged
over the reactor core volume is
depicted in Figure~\ref{coreneutron}.
The integrated 
flux is $7.6 \times 10^{13}~\nufluxunit $,
with 26\%, 52\% and 22\% in the 
thermal ($<$1~eV), 
epithermal (1~eV to 1 MeV) 
and fast ($>$1~MeV) ranges, respectively. 
The maximal flux at the center of the reactor core
is about 2.5~times the average value.

Comparisons were made
between these results 
with industry-standard calculations and
actual  reactor operation data.
Agreement to within 10\% was achieved.
In particular, the important 
neutron capture process 
$\ung$ leads to the
accumulation of $^{239}$Pu
via $\beta$-decays of $^{239}$U.
The yield of $\rm{Y_n}$=0.59 per fission
agrees with the results from
an independent study\cite{russian}.
These consistency requirements 
posed constraints to possible systematic
effects. The leading uncertainties 
are expected to arise from
the modeling of the reactor core compositions,
and were estimated to be $<$20\%. 

The process $\ung$
generates two $\nuebar$'s 
from $\beta$-decays
of $^{239}$U. Adding to the 
6 $\nuebar$/fission
from the fission fragments, the 
total $\nuebar$-yield is 
therefore $\ynubar$=7.2~$\nuebar$/fission,
such that 
$\ynu / \ynubar \sim 1.6 \times 10^{-4}$.
In particular,
the $\nue$-e to $\nuebar$-e
event rate ratio
at the electron recoil
energy range of 300 to 750 keV
is $\sim 2 \times 10^{-4}$,
too small
to account for the factor of two excess
over the Standard Model values
in the measured $\nuebar$-e rates
recently reported by the MUNU experiment\cite{munu}.

\subsection{B. Studies of Intrinsic Neutrino Properties}

A high-purity germanium detector has
collected data with a
trigger threshold of 5~keV at a distance
of 28~m from the core 
at KS Plant.
Background at the range of 
1/(kg-keV-day) was achieved\cite{prl03},
comparable with those from
underground Dark Matter experiments.
These unique low energy data provide 
an opportunity to study directly the
possible anomalous effects from
reactor $\nue$. 
Previous reactor experiments were
sensitive only to processes above 
the MeV range.
While the sensitivities
are not competitive to those
from reactor $\nuebar$\cite{munu,prl03}, 
the studies provide direct probes on 
the $\nue$ properties
without assuming CPT invariance, and
cover possible anomalous matter effects 
which may differentiate $\nue$ from $\nuebar$.

The anomalous coupling of neutrinos with 
photons are consequences of finite neutrino masses and
electromagnetic form factors\cite{nu04review}.
The manifestations
include neutrino magnetic moments ($\munu$) 
and radiative decays ($\Gamma_{\nu}$).
The searches of $\munu$ 
are usually performed in
neutrino-electron scattering experiments
$\rm{
 \nu_{\it l_1}   +    e^-   \rightarrow   \nu_{\it l_2}   +   e^-  .
  }$
Both diagonal and transition
moments are allowed, corresponding to the cases where
$l_1=l_2$ and $l_1 \neq l_2$, respectively.
The experimental observable is the kinetic energy of the
recoil electrons (T).
A finite neutrino magnetic moment ($\mu_l$),
usually expressed in units of the Bohr magneton
\begin{equation}
\rm{
\mub =  \frac{e}{2 m_e}  ~~ ; ~~ e^2 = 4 \pi \alpha_{em}  
}
\end{equation}
will contribute to a differential cross-section term
given by\cite{vogelengel}:
\begin{equation} 
\label{eq::mm} 
\rm{ 
( \frac{ d \sigma }{ dT } ) _{\mu}  ~ = ~
\frac{ \pi \alpha _{em} ^2 {\it \mu_l} ^2 }{ m_e^2 }
 [ \frac{ 1 - T/E_{\nu} }{T} ] ~
}
\end{equation}
where $\rm{\alpha_{em}}$ is the fine-structure constant,
$\enu$ is the neutrino energy and the natural unit
with $\hbar$=c=1 is adopted.
The quantity $\mu_l$ is an effective parameter which
can be expressed as\cite{beacom}:
\begin{equation}
\rm{
\mu_{\it{l}} ^2  =  \sum_j \big| \sum_k U_{\it{l}\rm{k}}  
\cdot \mu_{jk} \big| ^2  ~,
}
\end{equation}
where U is the mixing matrix and $\rm{\mu_{jk}}$
are the coupling constants
between the mass eigenstates
$\rm{\nu_j}$ and $\rm{\nu_k}$ with the photon.
Experimental signatures of $\mu_l$
from  reactor neutrino experiments
are therefore an excess of events between
reactor ON/OFF periods with an 1/T distribution. 

The 18-month reactor cycle suggests
that the optimal $\nue$'s are from  $\chr51$,
where the half-life is
$\tau_{\frac{1}{2}}$=27.7~days.
The equilibrium flux at 28~m is
$7.3 \times 10^{8} ~ \nufluxunit$.
With the actual reactor OFF period
denoted by t=0 to t=67~days,
the background-measuring $\offprime$ period
was taken to be from t=30 to 101 days, during
which the average residual $\nue$-flux
was 37\% of the steady-state ON-level.
The ON periods included data prior to reactor
OFF and starting from t=101~days.
A total of 3458/1445~hours of data from
the ON/$\offprime$ periods
were used in the analysis reported in this article.

\begin{figure}[htb]
\includegraphics[width=8cm]{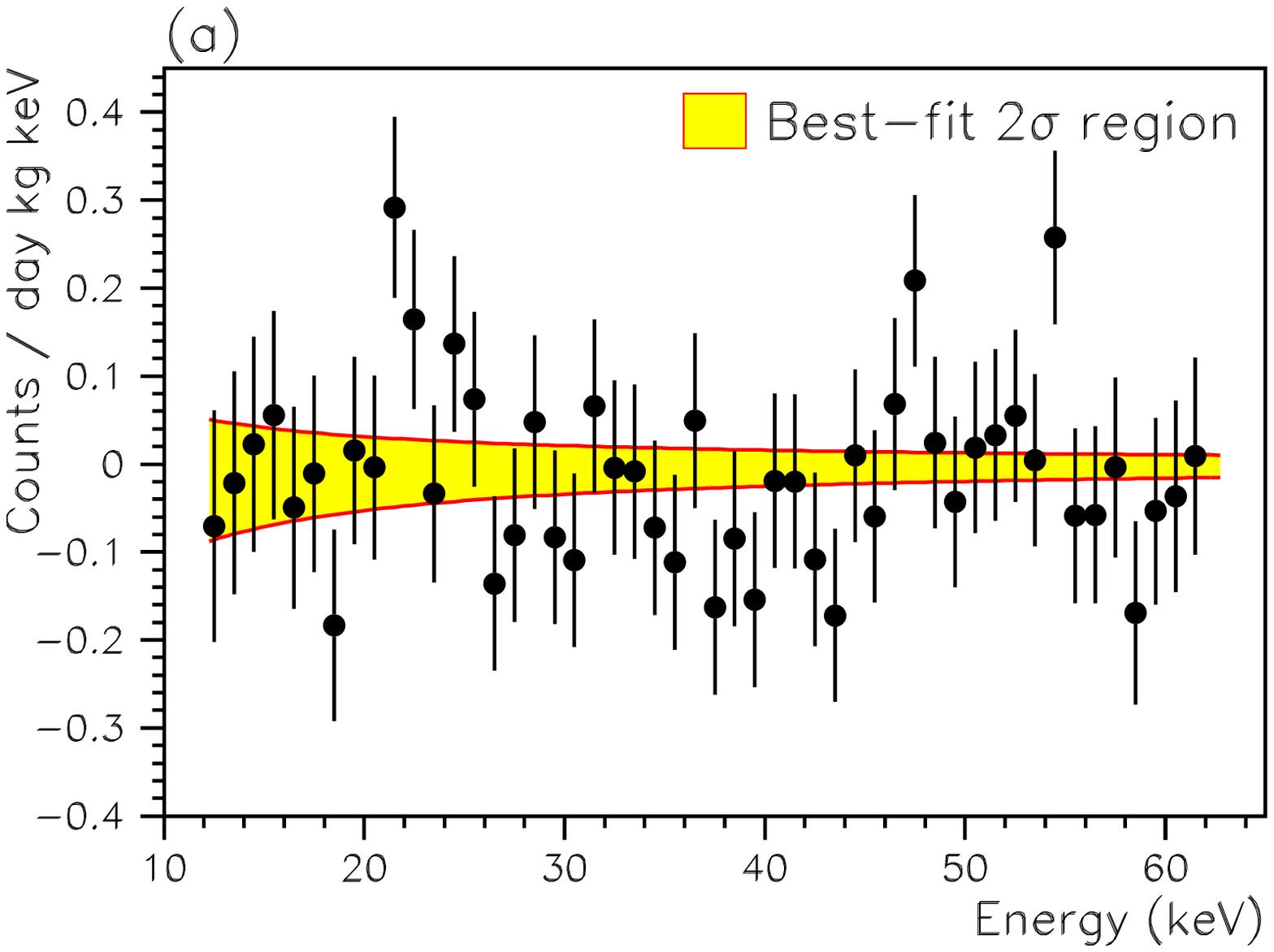}
\includegraphics[width=8cm]{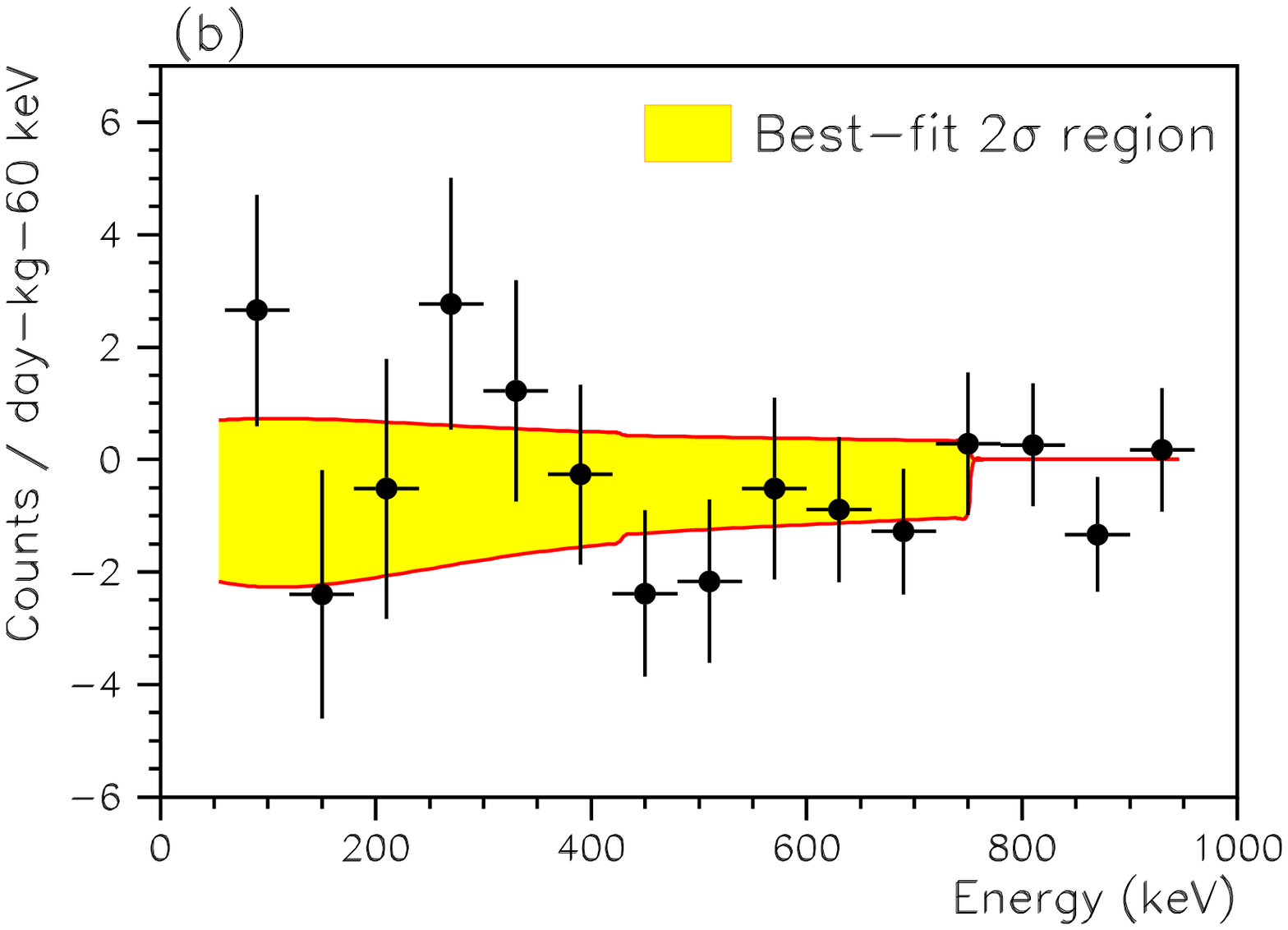}
\caption{
Residual plots for neutrino (a) magnetic moment
and (b) radiative decay searches
with the reactor $\chr51$ $\nue$-source.
}
\label{munu}
\end{figure}

The focus in the $\munu$-search
was on the T=10-100~keV range for
the enhanced signal rates and robustness
in the control of systematic uncertainties.
The $\nue$-e scattering
rates due to $\munu$ 
at the sensitivity level being explored 
are much larger (factor of 20 at 10~keV)
than the Standard Model rates from $\nuebar$,
such that the uncertainties in the irreducible
background can be neglected\cite{lernu}.
Similar event
selection and analysis procedures as
Ref.~\cite{prl03} were adopted.
Neutrino-induced events inside the Ge target
would manifest as ``lone-events'' uncorrelated
with the cosmic-ray veto panels
and the NaI(Tl) anti-Compton scintillators.
Additional pulse shape analysis 
further suppressed background due to electronic noise
and the delayed ``cascade'' events.
No excess of lone-events was observed in
the ON$-$$\offprime$ residual spectrum.  
A limit of
\begin{displaymath}
\munu  < 1.3 \times 10^{-8} ~ \mub
\end{displaymath}
at 90\% confidence level(CL) was derived.
The residual plot and the best-fit regions
are displayed in Figure~\ref{munu}a.

The neutrino-photon couplings
probed by $\nu$-e scatterings
can also give rise to
neutrino radiative decays:
$\rm{
\nu_j  \rightarrow   \nu_k   +   \gamma
}$
between mass eigenstates 
$\rm{\nu_j}$ and $\rm{\nu_k}$
with masses $\rm{m_j}$ and $\rm{m_k}$,
respectively.
The decay rates $\rm{\Gamma_{jk}}$
and half-lives $\rm{\tau_{jk}}$
are related to $\rm{\mu _{jk}}$ 
via\cite{rdk}
\begin{equation}
\label{eq::rdk}
\rm{
\frac{1}{\tau_{jk}} ~ = ~
\Gamma_{jk} ~ = ~
\frac{ \mu_{jk}^2 }{8 \pi} 
\frac{( m_j^2 - m_k^2 ) ^ 3 }{  m_j^3 } ~ .
}
\end{equation}

Results from oscillation experiments\cite{nu04,globalfit}
indicate that $\nue$ is predominantly a linear
combination of mass eigenstates $\nu_1$ and $\nu_2$
with mixing angle $\theta_{12}$ given by
$\rm{sin ^2 \theta_{12} \sim 0.27}$.
The mass differences between the mass eigenstates
are 
$\rm{\Delta m_{12} ^2 \sim 8 \times 10^{-5} ~ eV^2}$
and 
$\rm{\Delta m_{23} ^2 \sim 2 \times 10^{-3} ~ eV^2}$.
Both ``normal'' ($nor.$: $\rm{m_3 \gg  m_2 > m_1}$) and 
``inverted'' ($inv.$:  $\rm{m_2 > m_1 \gg  m_3}$) 
mass hierarchies are allowed.
The $\nu_1 \rightarrow \nu_3$ and
$\nu_2 \rightarrow \nu_3$ decays
are allowed only in the inverted
mass hierarchy, while
$\nu_2 \rightarrow \nu_1$
is possible in both hierarchies.
Adopting these as input,
the $\munu$ limit can be translated via
Eq.~\ref{eq::rdk} to  indirect bounds of
\begin{eqnarray*}
\frac{\tau_{13}}{m_{1}^3} ( inv.: \nu_1 \rightarrow \nu_3 ) 
& > & 1 \times 10 ^{23} ~  s / eV^3
\\
\frac{\tau_{23}}{m_{2}^3} ( inv.: \nu_2 \rightarrow \nu_3 ) 
& > & 4 \times 10 ^{22} ~  s / eV^3
\\
\frac{\tau_{21}}{m_{2}^3} ( nor.+inv.: \nu_2 \rightarrow \nu_1 ) 
& > & 6 \times 10 ^{26} ~  s / eV^3
\end{eqnarray*}
at 90\% CL.
These limits are sensitive to the 
bare neutrino-photon
couplings and are therefore 
valid for neutrino radiative decays of
in vacuum.

It is also of interest to perform a 
direct search of 
$\rm{ 
\nu_e  \rightarrow   \nu_X   +   \gamma
}$
the signature of which
is a step-function convoluted with
detector efficiencies where 
the end-point is at 
$\enu = 753$~keV for $\nue$'s from
$\chr51$\cite{venice}.
As shown in Figure~\ref{munu}b, 
no excess of uncorrelated
lone-events 
was observed in the
residual spectrum from the
ON$- \offprime$ data.
A limit of 
\begin{displaymath}
\rm{ 
\tau_{\nu} / m_{\nu} > 0.11 ~ s / eV
}
\end{displaymath}
for $\nue$
at 90\% CL was derived. 
This implies 
\begin{eqnarray*}
\frac{\tau_{1}}{m_{1}} & > & 0.08  ~  s / eV
\\
\frac{\tau_{2}}{m_{2}} & > & 0.03  ~  s / eV
\end{eqnarray*}
in the mass eigenstate basis.
These direct radiative decay limits
apply to all the kinematically allowed
decay channels and 
cover possible 
anomalous neutrino radiative decay mechanisms in matter,
since the decay vertices are within the active
detector volume.
In particular, the matter-induced 
radiative decay rates
can be enhanced by a huge 
factor($\sim$$10^{23}$)\cite{matter}
in the minimally-extended model.

Previous
accelerator experiments
provided the other direct ``laboratory''
limits on the $\nue$ magnetic moments and radiative decay
rates: $\munu  < 1.1 \times 10^{-9} ~ \mub$\cite{munuacc} 
and
$\rm{ \tau_{\nu} / m_{\nu} > 6.4 ~ s / eV}$\cite{rdkacc},
both at 90\% CL.
The new limits from reactor $\nue$ 
are complementary to
these more stringent results,
since they probe parameter space
with lower neutrino energy
and denser target density which
may favor anomalous matter effects.
Astrophysical arguments\cite{raffeltbook}
placed bounds which are orders of magnitude stronger\cite{pdg},
but there are model dependence and implicit assumptions
on the neutrino properties involved\cite{nu04review}.
Limits were also derived from solar neutrinos,
through the absence of spectral distortion
in the Super-Kamiokande spectra:
$\munu < 1.1 \times 10^{-10} ~ \mub$
\cite{beacom,skmunu},
and the observational limits of solar
X- and $\gamma$-rays:
$\rm{ \tau_{\nu} / m_{\nu} > 7 \times 10^{9} ~ s / eV}$
\cite{solarrdk}, both at 90\% CL.
However, the compositions of the mass eigenstates being
probed are different from those due to  
$\nue$ flavor eigenstate at the production site
studied by the reactor and
accelerator-based experiments,
such that the interpretations of the limits
are not identical.

\begin{figure}[htb]
\includegraphics[width=8.0cm]{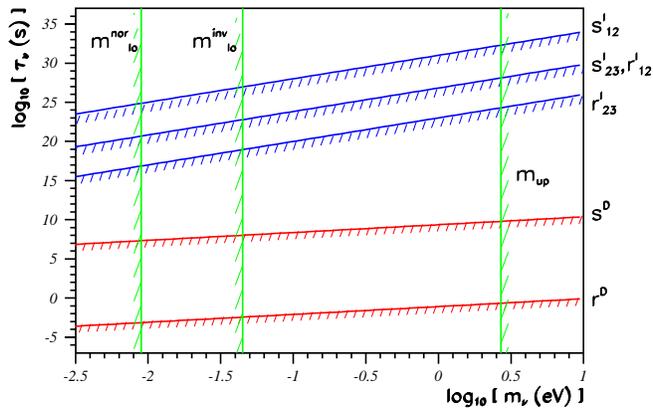}
\caption{
Summary of the results on
neutrino radiative
lifetimes for
$\nu_1$ and $\nu_2$
from reactor $\nue$ and solar
neutrinos experiments, denoted by
r and s, respectively. 
The superscripts (I,D) correspond  to
indirect bounds and direct limits,
while the subscript ``12'' is attributed
to decays driven by $\rm{\Delta m_{12}^2}$,
and so on. 
The upper bound ($\rm{m_{up}}$)
on $\rm{m_{\nu}}$ is due to limits
from direct mass measurements, while
the lower bounds
$\rm{m^{nor}_{lo}}$ and $\rm{m^{inv}_{lo}}$
are valid for 
the normal and inverted hierarchies,
respectively. 
The indirect bounds
$\rm{r^I_{23}}$ and $\rm{s^I_{23}}$ are valid
for inverted hierarchy only, while 
$\rm{r^I_{12}}$ and $\rm{s^I_{12}}$ 
apply to $\nu_2 \rightarrow \nu_1$ decays
in both hierarchies.
All modes are valid for decays in vacuum, while
$\rm{r^D}$ applies also for decays 
in matter. 
Bounds for $\nu_1$ and $\nu_2$
can be represented 
by the same bands in this scale.
}
\label{rdkplot}
\end{figure}

The limits for
the radiative decay lifetimes for
mass eigenstates $\nu_1$ and $\nu_2$ from reactor
and solar neutrino experiments
are summarized and depicted in Figure~\ref{rdkplot},
using the latest results from the neutrino 
experiments\cite{nu04,globalfit}
as input.
The notations are defined in the figure caption.
Several characteristic features can be identified.
The solar neutrino experiments lead to
tighter limits than those from
reactor $\nue$'s.
The indirect bounds inferred
from $\nue$-e scatterings are much more
stringent than the direct approaches,
but only apply to decays in vacuum.
Among the various approaches, only
the direct limit with reactor $\nue$ reported
in this article covers decays in both
vacuum and matter.


\section{III. Loaded Power Reactor}

\subsection{A. Enhancement of Neutrino Flux}

Using the simulation software discussed above,
we investigated  
the merits of inserting 
selected materials to the reactor core to enhance the
$\nue$-flux. 
A convenient procedure is to
load them to the unfilled rods or to
replace part of the UO$_2$ fuel elements 
or control rod assembly during reactor outage.
Though such a scenario involves 
difficulties with 
reactor operation regulations
and requires further 
radiation safety studies,
it is nevertheless technically feasible and
ready $-$
and costs much less than the
various accelerator neutrino factories projects,
which involve 
conventional neutrino beam upgrades\cite{superbeam}, 
muon storage rings\cite{muonring} and
beta beams\cite{betabeam}.
It is therefore of interest to explore 
the physics potentials and achievable sensitivities.

\begin{figure}[htb]
\includegraphics[width=8cm]{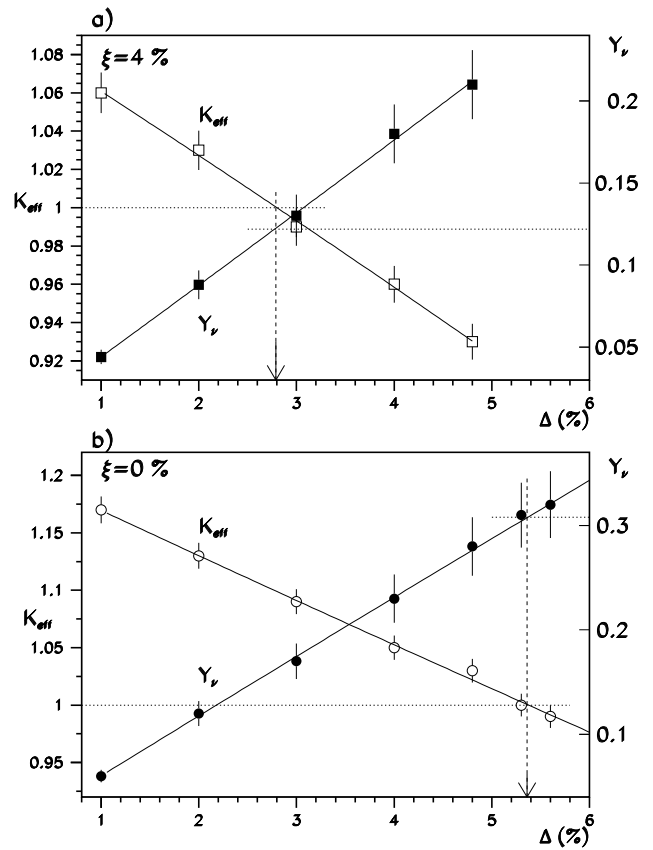}\\
\caption{
The variation of the  $\keff$ parameter and
the neutrino yield $\ynu$ from $^{50}$Cr(p) as
a function of loading fraction $\Delta$,
at control rod fraction (a) $\xi$=4\% and 
(b) $\xi$=0\%.
}
\label{loading}
\end{figure}

The candidate isotopes are those with good IA,
$\sigmang$ and BR as well as convenient lifetimes
for the activated nuclei. 
In order to sustain
the fission chain reactions (that is, 
having $\keff$=1),
the control rod fraction $\xi$ in the core
should be reduced and there is
a maximum amount of the neutron-absorbing materials
that can be inserted.
This amount as a
fraction of the fuel-element mass is
denoted by $\Delta$. 
Several selected materials and 
their maximal $\Delta$, $\rm{Y_n}$ and $\ynu$
at $\keff$=1 and $\xi$=0 in
both natural(n) and pure(p) IA
form are given in Table~\ref{artsource}.
The optimal choice is $^{50}$Cr(p).
To illustrate how
the allowed amount of control
rods and source materials 
would relate to the reactor operation,
the variations of $\keff$ and $\ynu$ versus $\Delta$ 
are plotted in Figures~\ref{loading}a\&b, at
two configurations where the control rod fractions
are (a) $\xi$=4\% and (b) $\xi$=0\%, respectively.
Criticality condition $\keff$=1
requires a maximum
load of  $^{50}$Cr(p) corresponding to
$\Delta$=5.4\% when the control rods
are completely retrieved ($\xi$=0\%).
This gives rise to 
a neutrino yield of $\ynu$=0.31~$\nue$/fission,
and therefore a $\ynu$/$\ynubar$ ratio of 0.04.
As shown in Table~\ref{mcnpinput},
the total weight of non-fuel materials 
inside the containment vessel
is 1.14 times that of the fuel elements.
Therefore, such loading of $^{50}$Cr(p) is only a
small addition of materials to the reactor core.
In the case of a $\rm{P_{th}}$=4.5~GW reactor,  
this maximal loading implies
8900~kg of $^{50}$Cr(p). 
A total of $\rm{4.2 \times 10^{19}}$ of $\nue$'s
per second
are emitted from the core, equivalent to
the activity of a 1.1~GCi source.
The $\nue$-flux at 10~m
is $3.3 \times 10^{12} ~ \nufluxunit$.

\begin{table}
\caption{\label{artsource}
The (n,$\gamma$) and $\nue$-yields for
selected materials loaded to the reactor core,
at $\keff$=1 and $\xi$=0\%. 
}
\begin{ruledtabular}
\begin{tabular}{lcccccccc}
Isotope & IA & $\sigmang$
& $\tau_{\frac{1}{2}}$ & Q & BR 
& $\Delta$ & $\rm{Y_n}$  & $\ynu$ \\ 
& (\%) & (b) & & (keV) & (\%) & (\%) & & \\ \hline
$^{50}$Cr(n) & 4.35 & 15.8 & 27.7~d  & 753 & 100 & 14.3 & 0.056 & 0.056 \\
$^{50}$Cr(p)  & 100 &  & & & & 5.4 & 0.31 & 0.31 \\ \hline
$^{63}$Cu(n) & 69.2 & 4.5 & 12.7~h & 1675 & 61 & 16.3 & 0.20 & 0.12 \\
$^{63}$Cu(p)  & 100 &  & & & & 14.8 & 0.25 & 0.15 \\ \hline
$^{151}$Eu(n) & 47.8 &  2800 & 9.3~h &  1920 & 27 & 0.073 & 0.092 & 0.025 \\
$^{151}$Eu(p) & 100 &  &  &  &  & 0.035 & 0.095 & 0.027
\end{tabular} 
\end{ruledtabular}
\end{table}

\subsection{B. Detection and Potential Applications}

In order to detect such neutrinos, 
detection mechanisms common to both $\nue$ and $\nuebar$
such as neutrino-electron scatterings are not appropriate.
Instead, flavor-specific charged-current interactions 
($\nue$NCC) would be ideal. 
Solar $\nue$ has been observed by $\nue$NCC in
radio-chemical experiments on
$^{37}$Cl and $^{71}$Ga, with
calibration measurements using $\chr51$ 
$\nue$-sources performed for $^{71}$Ga\cite{gasource}.
Detection of the low energy
solar neutrinos has been a central topic 
in neutrino physics. 
There are many detection schemes and 
intense research program 
towards counter experiments with $\nue$NCC\cite{lens}, 
using isotopes such as $^{100}$Mo, $^{115}$In, $^{176}$Yb.
The $\nue$NCC rates for the 
various isotopes in their natural abundance
at a $\chr51$ $\nue$-flux of
$3.3 \times 10^{12} ~ \nufluxunit$
are summarized in Table~\ref{nuNCC}. 
Also listed for 
comparison are the 
rates from the standard solar model 
$^7$Be $\nue$-flux
and from a 
1~MCi $^{51}$Cr source inside
a spherical detector of 1~m diameter.
More than $10^{4}$ $\nue$NCC events 
or 1\% statistical accuracy
can be achieved by one ton-year of
data with an indium target. 
Calculations of 
the $\nue$-flux 
depends on the amount of loaded
materials and the well-modeled reactor 
neutron spectra, so that
a few \% uncertainties should be possible.
Similar accuracies can be expected
on the $\nue$NCC cross-section measurements.
This would provide important calibration data
to complement the solar neutrino program.

\begin{table}
\caption{\label{nuNCC}
Expected $\nue$NCC rates
per ton-year at
a reactor $\chr51$ $\nue$-flux of
$3.3 \times 10^{12} \nufluxunit$($\rm{R_{core}}$),
at the standard solar model
$^7$Be flux($\rm{R_{\odot}}$), and
due to a 1~MCi
$\chr51$ source($\rm{R_{src}}$).
}
\begin{ruledtabular}
\begin{tabular}{lcccccc}
Target & IA(\%) & Threshold(keV)
& $\rm{R_{core}}$ &  $\rm{R_{\odot}}$
& $\rm{R_{src}} ^{\dagger}$ \\  \hline
$^{71}$Ga & 39.9 & 236 & 2100 & 3.8 & 58 \\
$^{100}$Mo & 9.63 & 168 & 2300 & 3.9 & 64 \\
$^{115}$In & 95.7 & 118 & 11000 & 19  & 300 \\
$^{176}$Yb & 12.7 & 301 & 3700 & 7.4 &  100
\end{tabular}
$^{\dagger}$ for four half-lives of data taking. \hfill
\end{ruledtabular}
\end{table}

Such mono-energetic $\nue$-sources
and the detection schemes may find applications
in other areas of neutrino physics.
We outline two of such applications
and derive their achievable statistical
accuracies. 
Discussions on the systematic
uncertainties and background of actual
experiments are beyond the scope of this
work, and will largely depend on the results
of the ongoing research efforts
to develop realistic $\nue$NCC-detectors.

The first potential application is on 
the study of the mixing angle $\theta _{13}$.
The mono-energetic $\nue$'s allow
simple counting experiments to be performed.
The rates between {\it NEAR} and {\it FAR} detectors 
can be compared 
to look for possible deviations
from 1/L$^2$, L being the core-detector distance.
The $\nue$-flux can be accurately measured
by the {\it NEAR} detectors, and the 
oscillation amplitude is precisely known
at fixed $\rm{\Delta m^2}$, $\enu$ and L.
Therefore, reactor $\nue$ experiments
are expected to have better
systematic control than those
with fission $\nuebar$'s\cite{theta13}
where, because of the 
continuous energy distribution,
the energy dependence in the
detector response and the oscillation effects
have to be taken into account.
Considering both oscillation and luminosity
effects, the sensitivities 
at a given neutrino energy $\enu$ 
depend on
$\rm{
[   sin^2 ( \frac{\Delta m^2 L}{ \enu } ) ] / \sqrt{L^2}
}$. 
The optimal distance for the 
{\it FAR} detector at
$\rm{\Delta m^2}$=0.002~eV$^2$
and $\enu$=747~keV for the $\chr51$-source
is therefore $\rm{L_{0}}=$340~m.
Table~\ref{theta13rate} shows
the achievable sensitivities to
$\rm{sin^2 2 \theta_{13}}$ 
with various detector options in
both natural(n) and pure(p) IA
located at $\rm{L_{0}}$ from a
two-core power plant each with 
$\rm{P_{th}}$=4.5~GW.
The source strength of
$\ynu$=0.31~$\nue$/fission 
for maximally-loaded $\chr51$
in Table~\ref{artsource} is adopted.
It can be seen that
a $\sim$1\% sensitivity 
can be statistically achieved with 5~years of 
data taking using a 100~ton indium
target $-$ a  
level comparable with those of 
the other 
reactor- and accelerator-based projects.

\begin{table}
\caption{\label{theta13rate}
Sensitivities to $\theta _{13}$
from maximally-loaded reactor core
with $\chr51$(p) sources for
different detector options.
Listed are
event rates per 500-ton-year(R$_{500}$)
for the {\it FAR} detector at  $\rm{L_{0}}=$340~m,
their achievable 1$\sigma$
statistical($\sigma _{500}$)
and
$\rm{ \sin ^2 2 \theta _{13} }$($\rm{\delta ( \sin ^2 2 \theta _{13} )}$)
accuracies.
}
\begin{ruledtabular}
\begin{tabular}{lcccc}
Target & IA(\%)
&  R$_{500}$
&  $\sigma _{500} (\%)$
& $\rm{\delta ( \sin ^2 2 \theta _{13} )}$\\ \hline
$^{100}$Mo(n) & 9.63  & 1900 & 2.3 & 0.027  \\
$^{100}$Mo(p) &  100  & 20000 & 0.71 & 0.0083 \\
$^{115}$In(n) & 95.7   & 9100 & 1.1 & 0.012 \\
$^{176}$Yb(n) & 12.7  & 3100 & 1.8 & 0.021 \\
$^{176}$Yb(p) &  100 & 24000 & 0.64 & 0.0075
\end{tabular}
\end{ruledtabular}
\end{table}

\begin{figure}[htb]
\includegraphics[width=8cm]{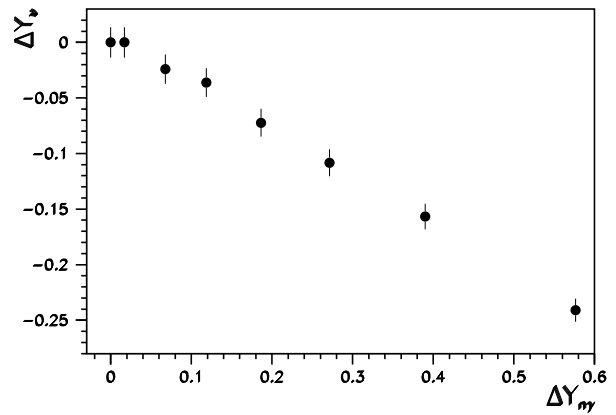}
\caption{
Simulated correlations between
the fractional changes of
$\nue$-yields ($\Delta \ynu$)
and those of
$\ur238$(n,$\gamma$)$^{239}$U rates
($\rm{\Delta Y_{n \gamma}}$)
in the case for a $\chr51$
source.
Conditions under which the
error bars are assigned
are explained in the
text.
}
\label{monitor}
\end{figure}

Another 
possibility is on the 
monitoring of 
unwarranted plutonium
production during
reactor operation $-$ an issue of paramount
importance in the control of nuclear 
proliferation\cite{ncontrol}.  
Plutonium is primarily produced by 
$\beta$-decays following
$\ung$$^{239}$U
whose cross-section  
is overwhelmed at high 
energy($>$1~eV)\cite{database}.
In contrast, the $\rm{( n , \gamma )}$
processes
in Table~\ref{artsource} which 
give rise to $\nue$-emissions are
predominantly thermal. 
The core neutron spectra 
can be modified without affecting the fission rates
through optimizations of the 
control rod and cooling water fractions,
making excessive plutonium production 
undetectable by monitoring the 
thermal power  output alone.
Measurements of the time-variations
of the $\nue$NCC event rates
are effective means to 
probe changes in the neutron spectra, and 
therefore to monitor directly the 
$\p239$ accumulation rates.
Illustrated in Figure~\ref{monitor}
are the correlations between
the fractional changes of
the $\nue$-yields ($\Delta \ynu$)
and those of the 
$\ur238$(n,$\gamma$)$^{239}$U rates
($\rm{\Delta Y_{n \gamma}}$)
in the case for a $\chr51$ source
having a strength of
$\ynu$=0.31~$\nue$/fission.
The uncertainties in $\Delta \ynu$  
correspond to those 
statistically achieved with
19~days of data
using a 10-ton indium detector located 
at 10~m from the reactor core.
Such a measurement is adequate to
make a 3$\sigma$ detection on
a 4\% reduction
of the $\nue$-flux, which corresponds to 
a 10\% enhancement of
the $\p239$ production rate.

\section{Acknowledgments}

The authors are grateful to
Drs. Z.L.~Luo and Y.Q~Shi
for
discussions on reactor neutron simulations.
This work was supported by contracts
91-2112-M-001-036 and 92-2112-M-001-057
from the National Science Council, Taiwan,
and 19975050 from the
National Science Foundation, China.

\end{document}